# BS-Patch: Constrained Bezier Parametric Patch

VACLAV SKALA, VIT ONDRACKA
Department of Computer Science and Engineering
University of West Bohemia
Univerzitni 8, CZ 306 14 Plzen
CZECH REPUBLIC
skala@kiv.zcu.cz　　http://www.VaclavSkala.eu

*Abstract:* - Bezier parametric patches are used in engineering practice quite often, especially in CAD/CAM systems oriented to mechanical design. In many cases quadrilateral meshes are used for tessellation of parameters domain. We propose a new modification of the Bezier cubic rectangular patch, the BS-patch, which is based on the requirement that diagonal curves must be of degree 3 instead of degree 6 as it is in the case of the Bezier patch. Theoretical derivation of conditions is presented and some experimental results as well. The BS-Patch is convenient for applications where for different tessellation of the $u-v$ domain different degrees of diagonal curves are not acceptable.

## 1 Introduction

Cubic parametric curves and surfaces are very often used for data interpolation or approximation. In the vast majority rectangular patches are used in engineering practice as they seem to be simple, easy to handle, compute and render (display). For rendering a rectangular patch is tessellated to triangles.

In this paper we describe a new cubic Bezier patch modification, called Bezier Smart-Patch (BS-patch). It is based on a Bezier cubic patch on which some additional requirements are applied. This modification is motivated by engineering applications, in general. It is expected that the proposed BS-patch can applied in GIS systems and geography applications as well.

## 2 Problem Formulation

Parametric cubic curves and surfaces are described in many publications [1]-[7], [9], [10], [14]. There are many different formulas for cubic curves and patches, e.g. Bezier, Hermite, B-spline etc., but generally diagonal curves of a cubic rectangular patch is a curve of degree 6. The proposed BS-patch, derived from the Bezier form, has diagonal curves of degree 3, i.e. curves for $v = u$ and $v = 1 - u$, while the original Bezier patch diagonal curves are of degree 6. Therefore the proposed BS-patch surface is "independent" of tessellation of regular $u-v$ domain. It means that if any tessellation is used, all curves, i.e. boundary and diagonal curves are of degree 3.

A cubic Bezier curve, see Fig.1, can be described in a matrix form as

$$p(t) = \boldsymbol{p}^T \boldsymbol{M}_B \boldsymbol{t}$$

$$\boldsymbol{M}_B = \begin{bmatrix} -1 & 3 & -3 & 1 \\ 3 & -6 & 3 & 0 \\ -3 & 3 & 0 & 0 \\ 1 & 0 & 0 & 0 \end{bmatrix} \quad (1)$$

where: $\boldsymbol{p} = [\,p_0,\ p_1, p_2, p_3]^T$ is a vector of "control" values of a Bezier cubic curve, $\boldsymbol{t} = [t^3,\ t^2,\ t,\ 1]^T$, $t \in \langle 0, 1 \rangle$ is a parameter of the curve and $\boldsymbol{M}_B$ is the Bezier matrix.

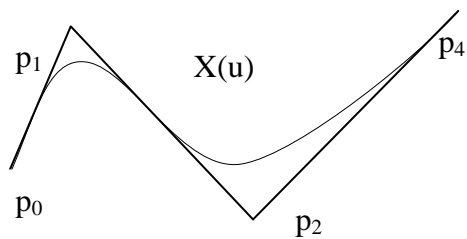

Fig.1 Bezier curve definition

A cubic Bezier patch, see Fig.2, is described in a matrix form for the $x$-coordinate as

$$x(u,v) = \boldsymbol{u}^T \boldsymbol{M}_B^T \boldsymbol{X} \boldsymbol{M}_B \boldsymbol{v} \quad (2)$$

where: $\boldsymbol{X}$ is a matrix of "control" values of the Bezier cubic patch

$$X = \begin{bmatrix} x_{00} & x_{01} & x_{02} & x_{03} \\ x_{10} & x_{11} & x_{12} & x_{13} \\ x_{20} & x_{21} & x_{22} & x_{23} \\ x_{30} & x_{31} & x_{32} & x_{33} \end{bmatrix} \quad (3)$$

$\boldsymbol{u}$, resp. $\boldsymbol{v}$, is a vector $\boldsymbol{u} = [u^3, u^2, u, 1]^T$, resp. $\boldsymbol{v} = [v^3, v^2, v, 1]^T$ and $u \in \langle 0,1 \rangle$, resp. $v \in \langle 0,1 \rangle$ is a parameters of the patch.

valid even theoretically the point lies on the plane).
2. The given $u - v$ rectangular domain mesh can be tessellated in different ways to a triangular mesh, in general, using different patterns, see fig.3.
3. If a rectangular Bezier cubic patch us used, then the diagonal curves, i.e. $v = u$ and $v = 1 - u$, are of degree 6.

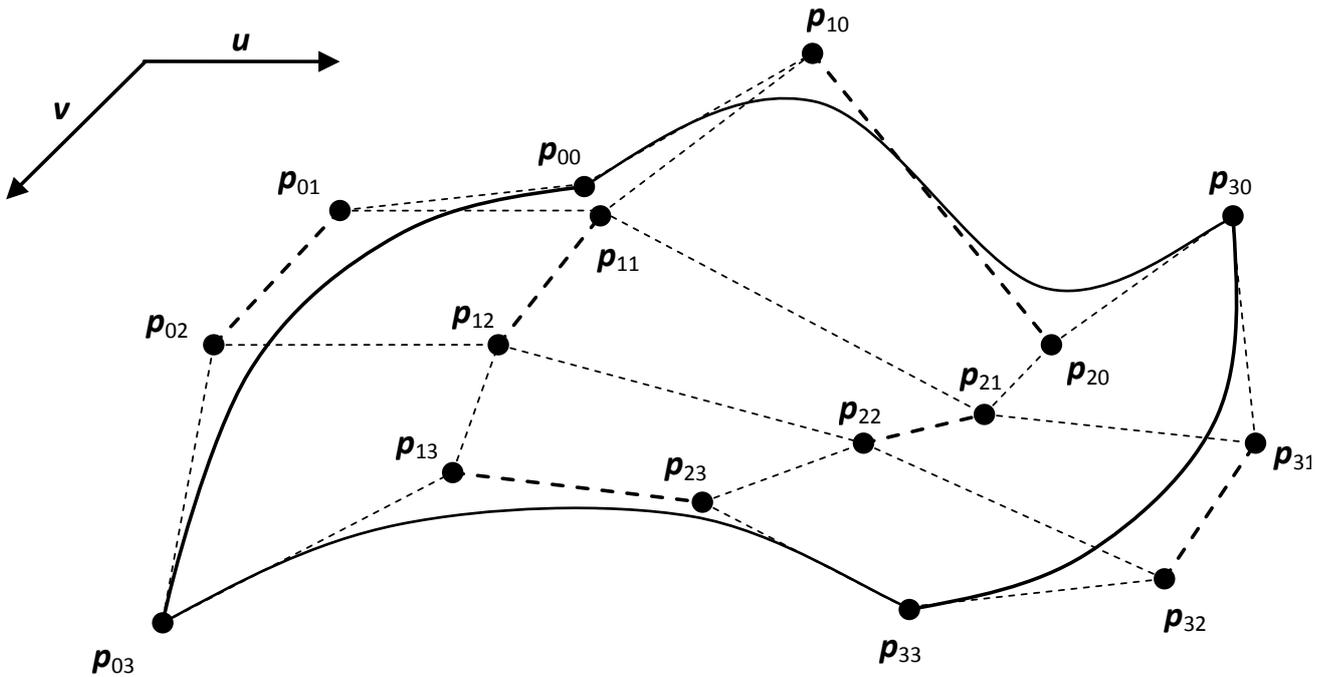

Fig.2 Control points of the Bezier patch

Similarly for $y$ and $z$ coordinates:

$$y(u,v) = \boldsymbol{u}^T \boldsymbol{M}_B^T \boldsymbol{Y} \boldsymbol{M}_B \boldsymbol{v}$$
$$z(u,v) = \boldsymbol{u}^T \boldsymbol{M}_B^T \boldsymbol{Z} \boldsymbol{M}_B \boldsymbol{v}. \quad (4)$$

It means that a rectangular Bezier patch is given by a matrix 4 x 4 of control values for each coordinate, i.e. by 3 x 16 = 48 values in E³.

From the definition of the Bezier patch it is clear, that boundary curves are cubic Bezier curves, i.e. curves of degree 3.

There are many applications, where a rectangular mesh is used in the $u - v$ domain. Sometimes $x$, resp. $y$ values are taken as $u$, resp. $v$ parameters and only $z$ value is interpolated/approximated as $y = f(x, y)$.

There are many reasons why patches, i.e. the $u - v$ domains, are tessellated to a triangular mesh, let us present just some of them:
1. A plane in E³ is defined by three points, so the 4th point is not generally on the plane (due to computer limited precision it is nearly always

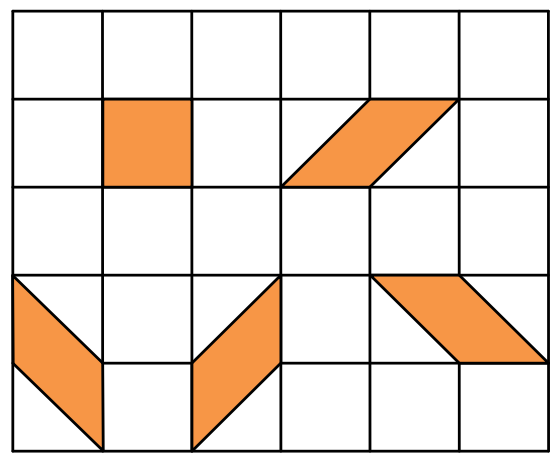

Fig.3 Different tessellation of $u - v$ domain for the given corner points

Especially last two points are very important as the final surface depends on tessellation and some curves might be of degree 6. This is not acceptable for some applications. There is a natural question:

*"Why some curves, i.e. when fixing $u = f(v)$, are degree of 3 and some are degree of 6?"*

If this feature is not controlled carefully it could lead to critical, sometimes even fatal, situations.

Understanding this, we exposed a specific restriction to the Bezier patch as curves for $v = u$ and $v = 1 - u$ must be of degree 3 as the patch boundary curves. This requirement has resulted into new modification of the Bezier cubic patch, called Bezier Smart-patch (BS-patch), described below. It follows development of the HS-Patch: Hermite Smart cubic patch modification [11]-[13].

## 3 Proposed BS-patch

Let us consider the Bezier patch on which we put some restrictions given by the requirement that diagonal curves, i.e. for $v = u$ and $= 1 - u$, are to be of degree 3.

The Bezier patch is given in the matrix form as

$$x(u,v) = \boldsymbol{u}^T \boldsymbol{M}_B^T \boldsymbol{X} \boldsymbol{M}_B \boldsymbol{v} \quad (5)$$

The restrictions for the proposed S-patch are:
- $x(u,v)$ for $v = u$ is a curve of degree 3, it means that $x(u) = \boldsymbol{u}^T \boldsymbol{M}_B^T \boldsymbol{X} \boldsymbol{M}_B \boldsymbol{u}$ is a curve of degree 3.

  We can write $(u) = \boldsymbol{u}^T \boldsymbol{R}_1 \boldsymbol{u}$,
  where $\boldsymbol{R}_1 = \boldsymbol{M}_B^T \boldsymbol{X} \boldsymbol{M}_B$

- $x(u,v)$ for $v = 1 - u$ is a curve of degree 3, it means that $x(u) = \boldsymbol{u}^T \boldsymbol{M}_B^T \boldsymbol{X} \boldsymbol{M}_B \boldsymbol{T} \boldsymbol{u}$ is a curve of degree 3,

where:

$$\boldsymbol{v} = [(1-u)^3 \quad (1-u)^2 \quad 1-u \quad 1]^T = \begin{bmatrix} -u^3 + 3u^2 - 3u + 1 \\ u^2 - 2u + 1 \\ -u + 1 \\ 1 \end{bmatrix} = \begin{bmatrix} -1 & 3 & -3 & 1 \\ 0 & 1 & -2 & 1 \\ 0 & 0 & -1 & 1 \\ 0 & 0 & 0 & 1 \end{bmatrix} \begin{bmatrix} u^3 \\ u^2 \\ u \\ 1 \end{bmatrix} = \boldsymbol{T} \boldsymbol{u} \quad (6)$$

We can write

$$x(u) = \boldsymbol{u}^T \boldsymbol{R}_2 \boldsymbol{u} \quad (7)$$

where

$$\boldsymbol{R}_2 = \boldsymbol{M}_B^T \boldsymbol{X} \boldsymbol{M}_B \boldsymbol{T} \quad (8)$$

The Bezier diagonal curve is in both cases defined as

$$\begin{aligned} x(u) &= \sum_{i,j=1}^{4} r_{ij} \, u^{4-i} \, u^{4-j} \\ &= r_{11}u^6 + (r_{12} + r_{21})u^5 \\ &+ (r_{13} + r_{22} + r_{31})u^4 \\ &+ (r_{14} + r_{23} + r_{32} + r_{41})u^3 \\ &+ (r_{24} + r_{33} + r_{42})u^2 \\ &+ (r_{34} + r_{43})u + r_{44} \end{aligned} \quad (9)$$

$$x(u) = \sum_{k=0}^{6} a_k \, u^k = \boldsymbol{a}^T \boldsymbol{u} \quad (10)$$

where: $\boldsymbol{u} = [u^6, \; u^5, \; u^4, \; u^3, \; u^2, \; u, \; 1]^T$

$$\boldsymbol{a} = [r_{11}, r_{12} + r_{21}, r_{13} + r_{22} + r_{31}, \\ r_{14} + r_{23} + r_{32} + r_{41}, r_{24} + r_{33} + r_{42}, \\ r_{34} + r_{43}, r_{44}]^T$$

We have 16 equations giving the relations between $x_{ij}$ and $r_{ij}$ for both cases, i.e. $\boldsymbol{R}_1$ and $\boldsymbol{R}_2$, that form a matrix relation, which expresses how the control values $x_{ij}$ form the coefficients in the $r_{ij}$

$$\boldsymbol{\rho} = \boldsymbol{\Omega} \, \boldsymbol{\xi} \quad (11)$$

where:

$$\boldsymbol{\rho} = [\, r_{11}, r_{12}, r_{13}, r_{14}, r_{21}, r_{22}, r_{23}, r_{24}, r_{31}, r_{32}, r_{33}, r_{34}, r_{41}, r_{42}, r_{43}, r_{44}\,]^T$$

$$\boldsymbol{\xi} = [\, x_{00}, x_{01}, x_{02}, x_{03}, x_{10}, x_{11}, x_{12}, x_{13}, x_{20}, x_{21}, x_{22}, x_{23}, x_{30}, x_{31}, x_{32}, x_{33}\,]^T$$

(12)

For the case 1, i.e. $u = v$ we get

$$\xi = [\ x_{00},\ x_{01},\ x_{02},\ x_{03},\ x_{10},\ x_{11},\ x_{12},\ x_{13},\ x_{20},\ x_{21},\ x_{22},\ x_{23},\ x_{30},\ x_{31},\ x_{32},\ x_{33}\ ]^T \quad (13)$$

$$\Omega_1 = \begin{bmatrix}
1 & -3 & 3 & -1 & -3 & 9 & -9 & 3 & 3 & -9 & 9 & -3 & -1 & 3 & -3 & 1 \\
-3 & 6 & -3 & . & 9 & -18 & 9 & . & -9 & 18 & -9 & . & 3 & -6 & 3 & . \\
3 & -3 & . & . & -9 & 9 & . & . & 9 & -9 & . & . & -3 & 3 & . & . \\
-1 & . & . & . & 3 & . & . & . & -3 & . & . & . & 1 & . & . & . \\
-3 & 9 & -9 & 3 & 6 & -18 & 18 & -6 & -3 & 9 & -9 & 3 & . & . & . & . \\
9 & -18 & 9 & . & -18 & 36 & -18 & . & 9 & -18 & 9 & . & . & . & . & . \\
-9 & 9 & . & . & 18 & -18 & . & . & -9 & 9 & . & . & . & . & . & . \\
3 & . & . & . & -6 & . & . & . & 3 & . & . & . & . & . & . & . \\
3 & -9 & 9 & -3 & -3 & 9 & -9 & 3 & . & . & . & . & . & . & . & . \\
-9 & 18 & -9 & . & 9 & -18 & 9 & . & . & . & . & . & . & . & . & . \\
9 & -9 & . & . & -9 & 9 & . & . & . & . & . & . & . & . & . & . \\
-3 & . & . & . & 3 & . & . & . & . & . & . & . & . & . & . & . \\
-1 & 3 & -3 & 1 & . & . & . & . & . & . & . & . & . & . & . & . \\
3 & -6 & 3 & . & . & . & . & . & . & . & . & . & . & . & . & . \\
-3 & 3 & . & . & . & . & . & . & . & . & . & . & . & . & . & . \\
1 & . & . & . & . & . & . & . & . & . & . & . & . & . & . & .
\end{bmatrix} \quad (14)$$

It should be read as e.g. $r_{42} = 3x_{00} - 6x_{01} + 3x_{02}$. Note that the notation used here is $\Omega_1\ \xi = r$

For the case 2, i.e. $v = 1 - u$ we get

$$\xi = [\ x_{00},\ x_{01},\ x_{02},\ x_{03},\ x_{10},\ x_{11},\ x_{12},\ x_{13},\ x_{20},\ x_{21},\ x_{22},\ x_{23},\ x_{30},\ x_{31},\ x_{32},\ x_{33}\ ] \quad (13)$$

$$\Omega_2 = \begin{bmatrix}
-1 & 3 & -3 & 1 & 3 & -9 & 9 & -3 & -3 & 9 & -9 & 3 & 1 & -3 & 3 & -1 \\
. & -3 & 6 & -3 & . & 9 & -18 & 9 & . & -9 & 18 & -9 & . & 3 & -6 & 3 \\
. & . & -3 & 3 & . & . & 9 & -9 & . & . & -9 & 9 & . & . & 3 & -3 \\
. & . & . & -1 & . & . & . & 3 & . & . & . & -3 & . & . & . & 1 \\
3 & -9 & 9 & -3 & -6 & 18 & -18 & 6 & 3 & -9 & 9 & -3 & . & . & . & . \\
. & 9 & -18 & 9 & . & -18 & 36 & -18 & . & 9 & -18 & 9 & . & . & . & . \\
. & . & 9 & -9 & . & . & -18 & 18 & . & . & 9 & -9 & . & . & . & . \\
. & . & . & 3 & . & . & . & -6 & . & . & . & 3 & . & . & . & . \\
-3 & 9 & -9 & 3 & 3 & -9 & 9 & -3 & . & . & . & . & . & . & . & . \\
. & -9 & 18 & -9 & . & 9 & -18 & 9 & . & . & . & . & . & . & . & . \\
. & . & -9 & 9 & . & . & 9 & -9 & . & . & . & . & . & . & . & . \\
. & . & . & -3 & . & . & . & 3 & . & . & . & . & . & . & . & . \\
1 & -3 & 3 & -1 & . & . & . & . & . & . & . & . & . & . & . & . \\
. & 3 & -6 & 3 & . & . & . & . & . & . & . & . & . & . & . & . \\
. & . & 3 & -3 & . & . & . & . & . & . & . & . & . & . & . & . \\
. & . & . & 1 & . & . & . & . & . & . & . & . & . & . & . & .
\end{bmatrix} \quad (14)$$

We can write

$$a_i = \sum_{j=1}^{16} \lambda_{ij} \xi_j = \lambda_i^T \xi \quad (15)$$

for $= 1, \ldots, 6$.

As we require the diagonal curves to be of degree 3, we can write conditions for that as:

- in the case 1: $r_{11} = 0$; $r_{12} + r_{21} = 0$; $r_{13} + r_{22} + r_{31} = 0$ using the matrix $R_1$

- in the case 2: $r_{11} = 0$; $r_{12} + r_{21} = 0$; $r_{13} + r_{22} + r_{31} = 0$ using the matrix $R_2$

From those conditions we get a system of linear equations

$$\Lambda\ \xi = 0 \quad (16)$$

where the first three rows of the matrix $\Lambda$ are taken for the case 1, i.e. related to the matrix $R_1$, and last three rows are taken for the case 2, i.e. related to the matrix $R_2$.

$$\xi = [x_{00}, x_{01}, x_{02}, x_{03}, x_{10}, x_{11}, x_{12}, x_{13}, x_{20}, x_{21}, x_{22}, x_{23}, x_{30}, x_{31}, x_{32}, x_{33}]^T \quad (17)$$

$$\Lambda = \begin{bmatrix} 1 & -3 & 3 & -1 & -3 & 9 & -9 & 3 & 3 & -9 & 9 & -3 & -1 & 3 & -3 & 1 \\ -6 & 15 & -12 & 3 & 15 & -36 & 27 & -6 & -12 & 27 & -18 & 3 & 3 & -6 & 3 & 0 \\ 15 & -30 & 18 & -3 & -30 & 54 & -27 & 3 & 18 & -27 & 9 & 0 & -3 & 3 & 0 & 0 \\ -1 & 3 & -3 & 1 & 3 & -9 & 9 & -3 & -3 & 9 & -9 & 3 & 1 & -3 & 3 & -1 \\ 3 & -12 & 15 & -6 & -6 & 27 & -36 & 15 & 3 & -18 & 27 & -12 & 0 & 3 & -6 & 3 \\ -3 & 18 & -30 & 15 & 3 & -27 & 54 & -30 & 0 & 9 & -27 & 18 & 0 & 0 & 3 & -3 \end{bmatrix} \quad (18)$$

The rank of the matrix $rank(\Lambda) = 5$, which means that we have to respect some restrictions generally imposed on the control points of the BS-patch. As the corner points are given by a user, the other control points are tied together with a relation. It can be seen that the vector $\xi$ is actually composed from values that are fixed (corner points are usually given) and by values, that can be considered as "free", but have to fulfill some additional condition(s). Let us explore this condition more in detail, now.

The equation $\Lambda \xi = 0$ can be rewritten as corner points are given as follows. Let us define vectors $\xi_1$ and $\xi_2$ as

$$\xi_1 = [x_{00}, x_{03}, x_{30}, x_{33}]^T$$

i.e. the corner points of the patch and

$$\xi_2 = [x_{01}, x_{02}, x_{10}, x_{11}, x_{12}, x_{13}, x_{20}, x_{21}, x_{22}, x_{23}, x_{31}, x_{32}]^T \quad (19)$$

i.e. other control points of the patch and matrices $\Lambda_1$ and $\Lambda_2$

$$\xi_1 = [x_{00}, x_{03}, x_{30}, x_{33}]^T$$

$$\Lambda_1 = \begin{bmatrix} 1 & -1 & -1 & 1 \\ -6 & 3 & 3 & 0 \\ 15 & -3 & -3 & 0 \\ -1 & 1 & 1 & -1 \\ 3 & -6 & 0 & 3 \\ -3 & 15 & 0 & -3 \end{bmatrix} \quad (20)$$

$$\xi_2 = [x_{01}, x_{02}, x_{10}, x_{11}, x_{12}, x_{13}, x_{20}, x_{21}, x_{22}, x_{23}, x_{31}, x_{32}]^T$$

$$\Lambda_2 = \begin{bmatrix} -3 & 3 & -3 & 9 & -9 & 3 & 3 & -9 & 9 & -3 & 3 & -3 \\ 15 & -12 & 15 & -36 & 27 & -6 & -12 & 27 & -18 & 3 & -6 & 3 \\ -30 & 18 & -30 & 54 & -27 & 3 & 18 & -27 & 9 & 0 & 3 & 0 \\ 3 & -3 & 3 & -9 & 9 & -3 & -3 & 9 & -9 & 3 & -3 & 3 \\ -12 & 15 & -6 & 27 & -36 & 15 & 3 & -18 & 27 & -12 & 3 & -6 \\ 18 & -30 & 3 & -27 & 54 & -30 & 0 & 9 & -27 & 18 & 0 & 3 \end{bmatrix} \quad (21)$$

It can be seen that we get

$$\Lambda_2 \xi_2 = -\Lambda_1 \xi_1 \quad (22)$$

that is equivalent to the equation $\Lambda \xi = 0$. Rewriting and reducing the system of equations above, we get

$$\begin{bmatrix} x_{01} & x_{02} & x_{10} & x_{20} & x_{11} & x_{12} & x_{21} & x_{22} & x_{13} & x_{23} & x_{31} & x_{32} \\ 1 & . & . & 2 & . & . & -3 & . & . & 1 & 2 & . \\ . & 1 & . & 1 & . & . & . & -3 & . & 2 & . & 2 \\ . & . & 1 & -2 & . & . & 3 & -3 & -1 & 2 & -3 & 3 \\ . & . & . & . & 1 & . & . & -2 & -1 & 2 & -1 & 2 \\ . & . & . & . & -1 & 1 & 1 & -1 & . & . & . & . \end{bmatrix} \xi_2$$

$$= \frac{1}{9} \begin{bmatrix} x_{00} & x_{03} & x_{30} & x_{33} \\ 6 & 3 & 12 & 6 \\ 3 & 6 & 6 & 12 \\ 3 & -3 & -12 & 12 \\ 2 & -2 & -2 & 11 \\ -1 & 1 & 1 & -1 \end{bmatrix} \xi_1 \quad (23)$$

From this equation we can see that the inner control points of the Bezier patch must fulfill the following condition:

$$x_{11} - x_{12} - x_{21} - x_{22} = \frac{1}{9}[x_{00} - x_{03} - x_{30} + x_{33}] \quad (24)$$

## 2.1 Constrain Conditions in the Hermite form - HS-Patch

Let us assume a Hermite bicubic patch, Fig.4. In [11] the following conditions for the Hermite form were derived the following conditions:

$$x_{33} + x_{44} = 2\varphi \quad x_{34} + x_{43} = 2\varphi$$
$$\varphi = x_{11} - x_{12} - x_{21} + x_{22} \quad (25)$$

where $\varphi$ is given by the corner points of the bicubic patch.

We can define two parameters $\alpha$ and $\beta$ (actually barycentric coordinates) as follows:

$$2\varphi\, \alpha = x_{44} \quad 2\varphi\,(1-\alpha) = x_{33} \quad (26)$$

and

$$2\varphi\, \beta = x_{43} \quad 2\varphi\,(1-\beta) = x_{34} \quad (27)$$

It means that the twist vectors are determined by $\alpha$ and $\beta$ values and by the corner points.

Now we have the following equations to be solved

1st row
$$b + x_{43} = b + 2\beta\varphi = -\varphi$$
$$b = x_{13} - x_{23} + x_{41} - x_{42} \quad (28)$$

2nd row
$$a + x_{44} = a + 2\alpha\varphi = -\varphi$$
$$a = x_{14} - x_{24} + x_{41} - x_{42} \quad (29)$$

3rd row
$$c - x_{43} - x_{44} = c - 2(\alpha + \beta)\varphi = -2\varphi$$
$$c = x_{31} - x_{32} - x_{41} + x_{42} \quad (30)$$

Then

$$\beta = -\frac{b+\varphi}{2\varphi} \qquad \alpha = -\frac{a+\varphi}{2\varphi} \quad (31)$$

$$-2\varphi = c - 2\varphi\alpha - 2\varphi\beta$$
$$= c + 2\varphi\frac{a+\varphi}{2\varphi} + 2\varphi\frac{b+\varphi}{2\varphi} \quad (32)$$

$$-2\varphi = c + a + \varphi + b + \varphi$$
$$a + b + c = -4\varphi \quad (33)$$

Expressing $\alpha$ and $\beta$ from the first two equations we get an equation (constraint) for the control values of the Hermite form that is actually the HS-patch as:

$$x_{31} - x_{32} + x_{41} - x_{42} + x_{14} - x_{24} - x_{23} + x_{13} = -4\varphi \quad (34)$$

i.e.

$$x_{31} - x_{32} + x_{41} - x_{42} + x_{14} - x_{24} - x_{23} + x_{13} = -4[x_{11} - x_{21} - x_{12} + x_{22}] \quad (35)$$

This result should be read as follows:
The Hermite HS-Patch control points are determined parametrically. The control points (tangent vectors) of the border curves have to fulfill the condition above. Twist vectors of the HS patch are controlled by values $\alpha$ and $\beta$ that are determined from control points that form the patch boundary.

## 2.2 Constrain Conditions in the Bezier form - BS-Patch

Unfortunately constrain conditions for the Bezier form, i.e. for BS-Patch, are not as simple as in the recently derived HS-Patch modification of the Hermite form [11], [12]. From the Hermite form we know that there are actually two parameters, i.e. analogous to $\alpha$ and $\beta$ in the HS-Patch, but their explicit formulation is just too complex.

It is easier to transform the Bezier form to the Hermite form and make all computations with the Hermite form, however the result is the same.

Mutual conversions between Hermite and Bezier patches are not generally presented in textbooks and can be derived. We present them below.

## 2.3 Hermite to Bezier conversion
In the following the Bezier control points will be determined by Hermite control points and vice versa.

$$\mathbf{X}_B = \begin{bmatrix} x_{00} & x_{01} & x_{02} & x_{03} \\ x_{10} & x_{11} & x_{12} & x_{13} \\ x_{20} & x_{21} & x_{22} & x_{23} \\ x_{30} & x_{31} & x_{32} & x_{33} \end{bmatrix}$$

$$= \begin{bmatrix} x_{11} & x_{11} + \frac{1}{3}x_{13} & x_{12} - \frac{1}{3}x_{14} & x_{12} \\ x_{11} + \frac{1}{3}x_{31} & x_{11} + \frac{1}{3}(x_{13} + x_{31}) - \frac{1}{9}x_{33} & x_{12} + \frac{1}{3}(x_{32} - x_{14}) - \frac{1}{9}x_{34} & x_{12} + \frac{1}{3}x_{32} \\ x_{21} - \frac{1}{3}x_{41} & x_{21} + \frac{1}{3}(x_{23} - x_{41}) - \frac{1}{9}x_{43} & x_{22} - \frac{1}{3}(x_{24} + x_{42}) - \frac{1}{9}x_{44} & x_{22} - \frac{1}{3}x_{42} \\ x_{21} & x_{21} + \frac{1}{3}x_{23} & x_{22} - \frac{1}{3}x_{24} & x_{22} \end{bmatrix} \quad (36)$$

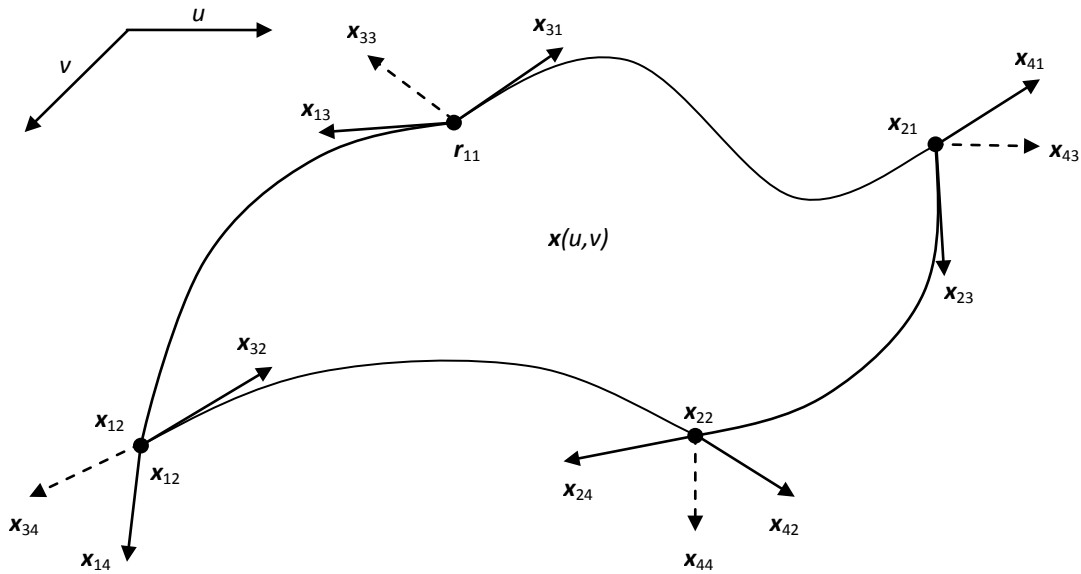

Fig.4 Control points of the Hermite patch

$$\boldsymbol{X}_H = \begin{bmatrix} x_{11} & x_{12} & x_{13} & x_{14} \\ x_{21} & x_{22} & x_{23} & x_{24} \\ x_{31} & x_{32} & x_{33} & x_{34} \\ x_{41} & x_{42} & x_{43} & x_{44} \end{bmatrix}$$
$$= \begin{bmatrix} x_{00} & x_{03} & 3(x_{01}-x_{00}) & 3(x_{03}-x_{02}) \\ x_{30} & x_{33} & 3(x_{31}-x_{300}) & 3(x_{33}-x_{32}) \\ 3(x_{10}-x_{00}) & 3(x_{13}-x_{03}) & 9(x_{00}-x_{01}-x_{10}+x_{11}) & 9(x_{02}-x_{03}-x_{12}+x_{13}) \\ 3(x_{30}-x_{20}) & 3(x_{33}-x_{23}) & 9(x_{20}-x_{21}-x_{30}+x_{31}) & 9(x_{22}-x_{23}-x_{32}+x_{33}) \end{bmatrix} \quad (37)$$

## 2.4 Bezier to Hermite conversion
The transformations above can be used to express similar constrain conditions for the Bezier form directly from the conditions for the Hermite form. However it can be seen that the conditions become very complex for a practical use.

## 4 Experimental Results
The experiments carried out proved that the proposed BS-patches and HS-patches have reasonable geometric properties.

To join BS-patches together a similar approach can be taken as for the standard Bezier patch in connecting patches, but there is just a small complication as the Eq (26) has to be respected and kept valid. It has just influence to a curvature of a surface of the neighbors.

To prove basic properties of the proposed BS-patch we used ½ of a cube. The approach proved that it is possible to join S-patches in a vertex and on edges smoothly, see Fig. 5.

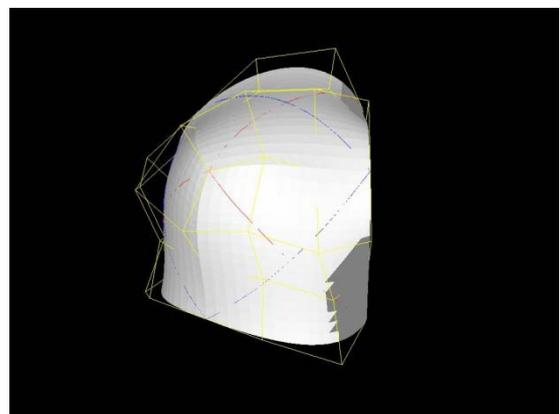

Fig.5 Joined patches

However, there was a severe problem detected, when the vertex of a mesh is shared by three patches. In some cases it was difficult to keep $C^1$ continuity, Fig.5.

Fig.6-8 presents the Utah Teapot modeled by the HS/BS Patches and rendered as HS-Patches. The edges of patches were highlighted to present the patches borders, the patches connections are $C^1$ or $G^1$ continuous similarly to standard patches connections. However it should be noted that the top of the cover there is only $C^0$ continuity and presents current semi limitation of the HS patch. It can be easily overcome by splitting the BS/HS patch horizontally into three patches.

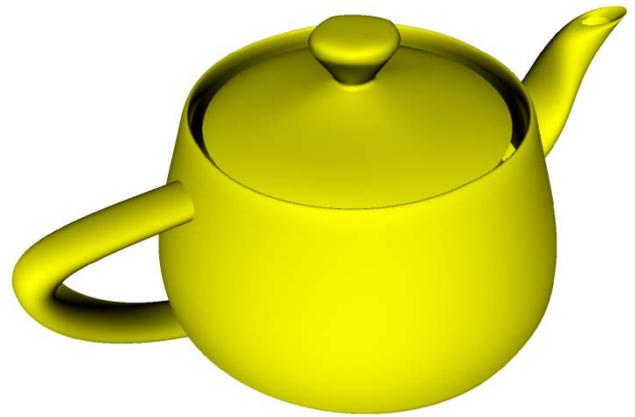

Fig.7 The Utah Teapot - HS Patch $C^1$ & $G^1$

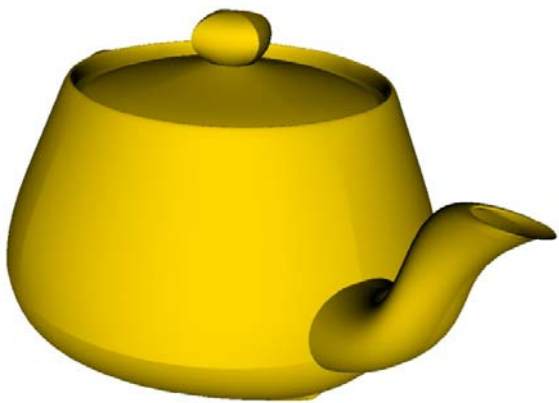

Fig.6 The Utah Teapot - HS Patch $C^1$ & $G^1$

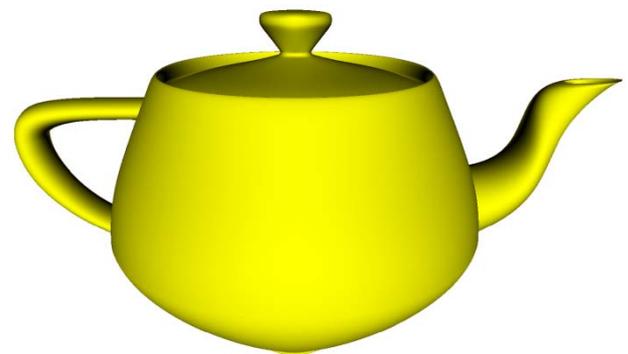

Fig.8 The Utah Teapot - HS Patch $C^1$ & $G^1$

Borders of patches are made explicitly visible as they have been rendered independently as a triangular mesh patch by patch. The surface is actually smooth.

At Fig.7 can be seen also violation of the condition for tangent vectors, while at Fig.8 and Fig.9 the top of the pot is split to more patches.

Fig.9 presents normal vectors of two joined patches for the HS-Patch which is identical also for the BS-Patch as well.

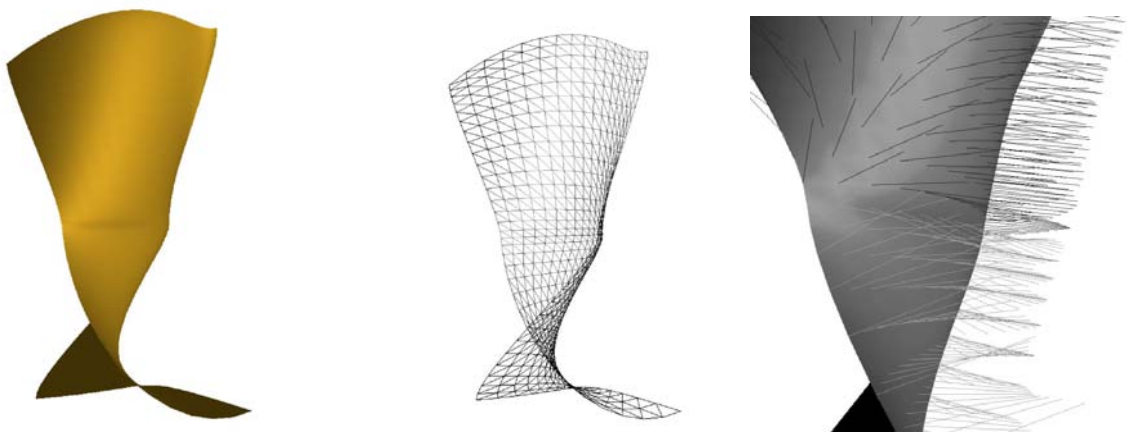

Fig.9 Two joined patches rendered, mesh and normal vectors

## 5 Conclusion

We have described and derived a new modification of the Bezier cubic patch. The main advantages of the proposed BS-patch are:

- Both diagonal curves are cubic curves, i.e. curves of degree 3.
- Different tessellations of $u - v$ domain and conversion to triangles do not change the degree of border and diagonal curves.
- Curves (boundary and diagonal) are of degree 3, less operations are needed as the computed polynomial is of degree 3.
- The given $u - v$ domain can be tessellated in different ways to four sided mesh and to triangular meshes for rendering using different tessellations.

It should be noted that one additional condition, i.e. Eq (25) must be kept valid, that a little bit complicates implementation, but on the other hand the presented advantages of the S-patch seems to be obvious.

Similar conditions were derived also for the B-Spline and Catmull-Rom patches and transformations are in detail described in [8].

## Future work

In experiments made several problems have been detected. However there is a one significant problem to be solved.

Let us imagine a situation when a corner is shared not by 4 patches, but by 3 patches only, e.g. if corner points form a cube and using 6 HS or BS-Patches we want to get object similar to a ball. In such cases vectors in corners are difficult to determine.

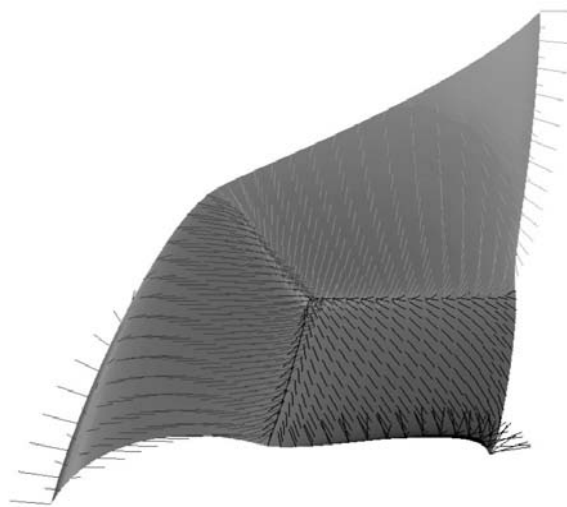

Fig.11 Normal vectors

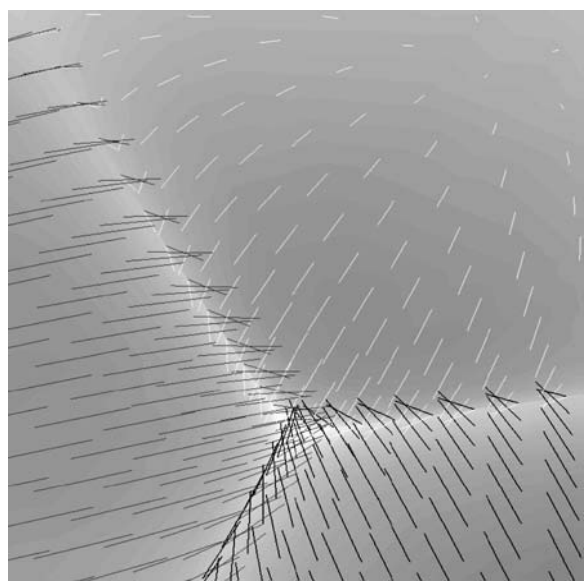

Fig.12 Normal vectors - detail

To join 3 patches having a common vertex seems to be complicated if the HS-Parch or BS-Patch conditions are to be fulfilled. This problem will be solved in the following research.

## Acknowledgment

The author would like to express to colleagues at the Center of Computer Graphics and Visualization for challenging this work, for many comments and hints they made, especially to Michal Smolik and Lukas Karlicek for rendering the Utah Teapot.

The project was supported by the Ministry of Education of the Czech Republic, projects NECPA No.LH12181 and VIRTUAL No.2C06002.

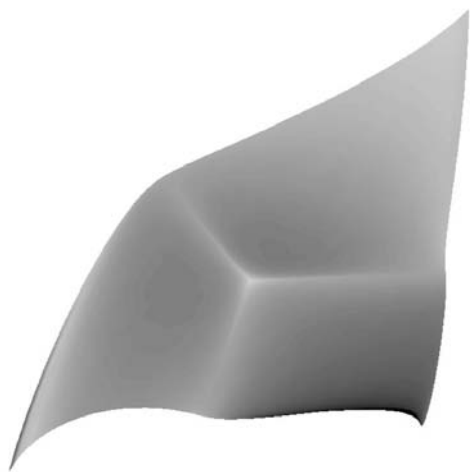

Fig.10 Joining 3 patches